\documentstyle[12pt]{article}
\def\d{\partial}

\def\Tr{{\rm Tr}}
\def\tr{{\rm tr}} 
\newcommand\0{\nonumber}
\newcommand\ee{\end{eqnarray}}	 	
\newcommand\be{\begin{eqnarray}}
\newcommand\ba{\begin{array}}			
\newcommand\ea{\end{array}}

\newcommand\nc{{noncommutative }}
\begin{document}
\begin{flushright}
{SISSA 74/01/EP}\\
{ hep-th/0109204}
\end{flushright}

\begin{center}
{\LARGE {\bf Chiral anomalies }}
\vskip .5cm

{\LARGE {\bf in noncommutative YM theories}} 
\vskip 1cm

{\large L. Bonora} 
{}~\\
\quad \\
{\em International School for Advanced Studies (SISSA/ISAS),}\\
{\em Via Beirut 2, 34014 Trieste, Italy and INFN, Sezione di Trieste}\\
 {\tt bonora@sissa.it}

\vskip .5cm
{\large A. Sorin}
{}~\\ 
\quad \\
{\em Bogoliubov Laboratory of Theoretical Physics,}\\
{\em Joint Institute for Nuclear Research (JINR),}
\\
{\em 141980, Dubna, Moscow Region, Russia}\\
{\tt sorin@thsun1.jinr.ru}
\end{center}

\vskip 2cm
{\bf Abstract.} Using cohomological methods we discuss several issues 
related to chiral anomalies in
 noncommutative $U(N)$ YM theories
in any even dimension. We show that for each dimension there is only one
solution of the WZ consistency condition and that there cannot be any
reducible anomaly, nor any mixed anomaly when the gauge group is a product
group. We also clarify some puzzling aspects of the issue of the
anomaly when chiral fermions are in the adjoint representation. 
  
\vskip 1cm 
 
\section{Introduction}

The subject of chiral anomalies in \nc $U(N)$ gauge field theories has been 
addressed by several authors \cite{Haya,Ard,GM,BST,Mor,Mar,Int}.
The generally accepted conclusion is that in 
order for  \nc gauge theories to be anomaly--free they must be nonchiral
(which includes also formally chiral theories with adjoint matter in
D=4) and that mixed anomalies are absent. In this paper we would
like to add some further evidence to these conclusions and extend them
to dimensions other than 4. 

The method we employ is based on 
the WZ consistency conditions \cite{WZ} and relies on the concept of
{\it nc-locality} (almost an oxymoron), which means that the space
of cochains we consider is the same as in ordinary local field theories
with the ordinary product replaced by the Weyl--Moyal product.
This principle of nc--locality is suggested by one--loop renormalization
of \nc field theories, where counterterms are precisely of the above 
type, and, in the cases in which the \nc field theories can be embedded
in string theory in the presence of a B field, can be traced back to 
the properties of  (tree and planar one--loop) string amplitudes, 
precisely to the fact that such 
 string amplitudes factorize into
\nc factors and ordinary string 
 amplitudes. However we do not know
whether nc-locality is compatible with IR--UR
and with higher loops
renormalizations. Therefore, for the time being we take it as a working
hypothesis. The advantage of using this 
 method is that,
once the formalism is established, many conclusions
 are evident without
resorting to explicit Feynman diagram calculations.
 
This paper is an elaboration upon \cite{BST}, where descent equations 
for anomalies in \nc theories were introduced. It is organized as 
follows.
In the next section we introduce our notation. In
section 3 we discuss the problem of deriving anomalies from the descent
equations and concentrate in particular on the uniqueness of the
solutions. Section 4 is devoted to the anomaly problem in the adjoint
representation. We show that in $D=4k$ this anomaly vanishes and in
$D=4k+2$ is equal to $2N$ times the anomaly in the fundamental 
representation. In section 5 we summarize our results.

\section{Notation and conventions}

In the following we will consider $U(N)$ gauge theories in a 
noncommutative ${\bf R}^D$, with Moyal deformation parameters 
$\theta^{\mu\nu}$. The gauge potential will be denoted by
$A_\mu^j{}_i$ with $i,j=1,\dots,N$ being the indices of the fundamental
and antifundamental representation of $U(N)$. Next we introduce a basis of
hermitean matrices 
 $t^A= (t^A)^j{}_i$, (capital letters $A,B,...=0,\ldots
N^2-1$ 
 will denote indices in the Lie algebra $u(N)$), with the
normalization  
 \be
{\rm tr} (t^A t^B)= \frac 12 \delta^{AB}~\label{trAB}.
\ee
This can be done, for example, by using a basis of hermitean matrices
for the Lie algebra of $SU(N)$, $t^a$, (whenever
necessary, lower case letters $a,b,...=1,\ldots N^2-1$ will denote 
indices in the adjoint of $su(N)$), and adjoining 
$t^0= \frac{1}{\sqrt{2N}} {\bf 1}_N$. The basis $t^A$ satisfies
\be
[t^A,t^B]= i f_{ABC}t^C, \quad \{t^A,t^B\}= d_{ABC}t^C\label{fd}
\ee
where $f_{ABC}$ is completely antisymmetric, $f_{abc}$ is the same as 
for $su(N)$ and $f_{0BC}=0$, while $d_{ABC}$ is completely symmetric; 
$d_{abc}$ is the same as for $su(N)$,
$d_{0BC}=\sqrt{\frac {2}{N}} \delta_{BC}$, $d_{00c}=0$ and $d_{000}=
\sqrt{\frac {2}{N}}$, see \cite{BS}. Here and henceforth summation over
repeated indices is understood and upper or lower indices are used
interchangeably
since the metric is $\delta_{AB}$.

Using this basis we write
\be
A_\mu^j{}_i \equiv A_\mu^B (t^B)^j{}_i,\quad\quad\quad A_\mu= 
A_\mu^B t^B, 
\quad\quad \quad A_\mu^B =2\, \tr(t^B A_\mu)\0
\ee
and $\tr$ denotes the trace in the fundamental representation.

With this notation, the action of the chiral fermions in the fundamental
representation interacting with an external gluon is
\be 
S = \int d^Dx \, \bar \psi\star \gamma^\mu(i \d_\mu \psi 
+ A_\mu \star P_+\psi) 
\label{actfund} 
\ee 
where $P_\pm = \frac 12 (1\pm \widehat \gamma)$ and  
$\widehat \gamma=i^{1-n}\gamma_0\gamma_1\ldots\gamma_{D-1}$, with $D=2n$.

For later use we introduce also the $N^2\times N^2$ matrices
$F^A, D^A$ 
 
\be
(F^A)^{BC} =f^{BAC},\quad\quad (D^A)^{BC}= d^{BAC} \label{FD}
\ee
They satisfy the commutation rules
\be
[F^A,F^B]=f^{ABC} F^C,\quad\quad [F^A,D^B]=f^{ABC} D^C,
\quad\quad [D^A,D^B]=-f^{ABC} F^C. \label{commFD}
\ee
Now let us form the combination $G^A=\frac 12 (D^A+i F^A)$. It is easy to
prove
 that they are hermitean and satisfy the relations (use the identities
in \cite{BS})
\be
[G^A,G^B]= i f^{ABC}G^C, \quad \{G^A,G^B\}= d^{ABC}G^C\label{G}
\ee
isomorphic to (\ref{fd}). We notice that, if $^T$ denotes 
transposition, then  
\be
[G^A, (G^B)^T]=0.\label{comm}
\ee
We also have 
\be
\widehat{\tr}(G^AG^B) = \frac N2 \delta^{AB}\label{TrAB}
\ee
$\widehat{\tr}$ is the trace in the representation of $u(N)$ spanned by the 
$G^A$'s, while the symbol $\Tr$ will be used in a generic sense without 
regard to a particular representation.
In the following we will need to consider the combinations
 \be
\widehat A_\mu= A_\mu^B G^B\label{hatA}.
\ee
One may wonder what is the representation of $u(N)$ spanned by the
generators $G^A$. This representation 
is equivalent to the direct sum of $N$ copies of the fundamental 
representation. A way to see this is by computing 
traces of generators. By repeatedly using (\ref{fd}) 
on one side and (\ref{G}) on the other side, and, finally, utilizing
(\ref{trAB}) and (\ref{TrAB}) one can easily show that
\be
\widehat{\tr}(G^{A_1}...G^{A_n})= N\,\tr(t^{A_1}...t^{A_n}).\label{trNTR}
\ee

\section{Anomalies from cocycles}

It is well--known by now that only the (anti)fundamental and the adjoint
representations of $u(N)$ extend to representations of the Lie algebra
of \nc $U(N)$ gauge transformations, \cite{shahin}. 
So we can build \nc gauge
theories
 only with the latter representations or direct sums of them.  

Taking this into account, let us give a more detailed reformulation of
the approach in \cite{BST}. To this end we consider a
one--form gauge potential ${\cal A}={\cal A}_\mu dx^\mu= A^B X^B$, 
with gauge field strength two--form 
${\cal F} = d{\cal A} +i {\cal A}\star {\cal A}=F^BX^B$ and
gauge 
 transformation parameter ${\cal C}=C^BX^B$ (which we take to be a
Grassmann--odd 
 Faddeev--Popov ghost with ghost number 1). All these
quantities are valued in the Lie algebra generated by $X^A$ which stand
either for $t^A$ or $G^A$'s or 
 by direct sums of them.
They are therefore hermitean matrices.
 
The gauge (BRST) transformations are:
\be
s{\cal A} = d{\cal C} - i {\cal A}\star {\cal C} + i{\cal C}\star {\cal A},
\quad\quad s{\cal C} = -\, {\cal C}\star {\cal C}\label{brst}
\ee
$d$ and $s$ are assumed to commute. As a consequence
the transformations (\ref{brst}) are nilpotent as in the ordinary case. 

Now, as in ordinary theories, we would like to write down the descent
equations
\cite{WZ,S,BC,ZWZ,Lang} relevant to $D=2n$ dimensions, starting
from a closed and
 BRST invariant $(2n+2)$--form $\Omega_{2n+2}$, constructed
as a polynomial 
 of ${\cal F}$ and referred to as the {\it top form}:
\be
&&\Omega_{2n+2}= d \Omega_{2n+1}^0\0\\
&&s \Omega_{2n+1}^0= d \Omega^1_{2n}\label{descent}\\
&&s \Omega^1_{2n} = d \Omega^2_{2n-1}\0
\ee
where the upper index is the ghost number and the lower index is the
form order. $\Omega^1_{2n}$ is the (unintegrated) anomaly. 
The virtue of the descent equations formalism is that it provides
explicit expressions for anomalies and one is spared the details of the 
complicated verification 
that $\Omega^1_{2n}$ does satisfy the Wess--Zumino consistency 
conditions. The latter is an automatic consequence of the top 
form $\Omega_{2n+2}$ being closed and invariant (and, of course, nontrivial,
i.e. non--exact, 
 otherwise the corresponding anomaly would be trivial). 

However in \nc gauge theories there is a complication. The above method 
does not work straightforwardly, because there exists no closed 
invariant polynomial that can be built with the \nc curvature ${\cal F}$.  
But there is a way out that was pointed out in \cite{BST}: 
the differential space of cochains must be constituded by forms that
are defined up to an overall cyclic permutations of the Moyal
product factors
involved. This will be spelt out in more detail in a moment
(see the definition of cyclic equivalence below).

But, before, let us pause to make a comment on this method. At first sight
it may look artificial, but it is a very effective method to
derive the expression of the anomaly. What is relevant is that the 
last equation in (\ref{descent}) can now be rewritten as
\be
s \Omega^1_{2n} = d \Omega^2_{2n-1}+\ldots \0
\ee
where dots denote terms that can be cast in the form of graded
$\star$--commutators. It is 
 well--known that these terms are total
derivatives of the form 
 $\theta^{\mu\nu}\d_\mu...$ . So upon integration,
this equation gives
 
\be
s \left(\int d^Dx \, \Omega^1_{2n}\right)=0
\ee
which precisely says that $\int d^Dx \, \Omega^1_{2n}$ satisfies the
Wess--Zumino consistency conditions.

From now on the descent equations (\ref{descent}) have to be understood
in the framework of the new definition of the BRST cohomology.

What the above discussion boils down to is that, in order to know
what anomalies we have in a given theory, we can simply concentrate
on the possible closed and invariant forms $\Omega_{D+2}$ we can 
construct out of the curvature ${\cal F}$. In \cite{BST} the forms considered
were simply traces of $\star$-products of $F$ in the fundamental 
representation. Here we wish to be more general. The Feynman rules tell
us that the anomaly will contain traces of the matrices that appear
in the fermion--gluon vertex, i.e. traces of $t^A$ or $G^A$ or direct 
sums of them (let us denote them collectively by $X^A$). 
However one cannot exclude a priori that these traces may have
symmetry properties in some of the indices (similarly to what happens 
in ordinary theories). Therefore we will limit ourselves to writing
the polynomials that are involved in the descent equations
(\ref{descent}) as
\be
h^{A_1A_2...A_k}E^{A_1}_1\star E^{A_2}_2\star 
\ldots \star E^{A_k}_k\label{poly}
\ee
where $E^{A_i}_i$ is any form of the type $C^{A_i}, A^{A_i}$ or 
exterior
differentials of them, and $h$ is a tensor obtained as
a combination of traces of the appropriate generators. All the polynomials
appearing in
 the descent equations are considered the same if they differ by
a cyclic ordering of the factors $E_1,...,E_k$ (with the correct 
grading). We refer to the latter identification as 
{\it cyclic equivalence}.

As for the forms $\Omega_{D+2}$, we will write them as
\be
h^{A_1A_2...A_{n+1}}F^{A_1}\star
F^{A_2}\star \ldots \star F^{A_{n+1}}.
\label{polymax}
\ee
Due to the cyclic equivalence, we can assume that the $h^{A_1\dots A_{n+1}}$
is cyclically symmetric. 
As pointed out above the forms (\ref{polymax}) must be closed and BRST
invariant. Due to the 
 Bianchi identity, these two requirements amount to the
same property.
Using
\be
\Gamma\star \Lambda= G^{ABC}\Gamma^A\star \Lambda^B X^C\0
\ee
for any two forms $\Gamma= \Gamma^AX^A$ and $\Lambda=\Lambda^AX^A$ and
cyclicity of $h$, one can see that the latter must satisfy
\be
h^{A_1...XC...A_{n+1}}G^{XBD}- h^{A_1...BX...A_{n+1}} G^{XDC}=0
\label{hconstr}
\ee
for any couple of contiguous indices $B,C$. This set of constraints
together with the cyclic symmetry in the indices characterize the
tensors $h$. 
 
It is perhaps useful to remark that in ordinary theories the condition
(\ref{hconstr}) is replaced by a weaker one, which therefore allows in
general for more solutions.

Now, let us examine the consequences of (\ref{hconstr}). 
The simplest case, n=1 (D=2), is trivial; we can only have
$h^{AB}\sim \delta^{AB}$, which correspond to the trace of the product
of two generators, and is easily seen to satisfy (\ref{hconstr}). 
The next case, n=2 (D=4), has two possibilities: either $h^{ABC}=
f^{ABC}$ or $h^{ABC}=d^{ABC}$. Using the identities for the $f,d$
tensors, see \cite{BS}, one can see that (\ref{hconstr}) is not
satisfied for either of these possibilities separately, while it 
is satisfied
for the combination $d^{ABC}+if^{ABC}$. But this precisely means
that $h^{ABC}\sim \Tr (X^AX^BX^C)$. One can similarly proceed to 
higher dimensions and convince oneself that the only solution is
in any case $h^{A_1...A_n}\sim\Tr (X^{A_1}\ldots X^{A_n})$. 

Therefore we end up with the ansatz made in \cite{BST} for the top form 
of $\Omega_{D+2}$, with the additional specification that ${\cal F}=F^AX^A$,
where $X^A$ can be $t^A, G^A$ or direct sums of them. Now, one has
simply to apply the formulas of \cite{BST} to get the anomalies in 
any even $D=2n$ dimension. It is the one determined by the top form
$\Omega_{D+2}=\tr({\cal F}\star\ldots \star{\cal F})$ with $n+1$ entries. 
The corresponding (unintegrated) anomaly is given  by the formula, 
see \cite{BST},
\be
\Omega^1_{2n} &=& 
n\int_0^1 dt (t-1)\Tr(d{\cal {\cal C}}\star 
{\cal A}\star {\cal F}_t\star ...\star {\cal F}_t
+d{\cal C}\star {\cal F}_t\star {\cal A}\star ...\star {\cal F}_t+\0\\
&&\ldots+d{\cal C}\star {\cal F}_t\star {\cal F}_t\star ...\star {\cal A}) 
 \label{anom}
\ee
where the sum under the trace symbol includes $n-1$ terms. In (\ref{anom})
we have introduced a parameter $t$, $0\leq t\leq 1$, and  the traditional
notation ${\cal F}_t= td{\cal A} +it^2{\cal A}\star {\cal A}$.

For example the anomaly of the action (\ref{actfund}) is obtained from the
above formula by replacing ${\cal A}, {\cal C}$ and ${\cal F}$ with
the corresponding fields in the fundamental representation
$A,C$ and $F$, respectively, by integrating the expression (\ref{anom})
over the space-time and multiplying it by the factor
\be
\frac {2^n}{(n+1) (4\pi)^n \Gamma (n+1)}.\0
\ee

Let us now draw some conclusions. 
The first is that in \nc gauge theories, as
opposed to ordinary ones, there cannot
be
 reducible anomalies, that is anomalies derived from a top form
made of product of traces
such as $\Tr({\cal F}\star ...\star {\cal F})
\Tr({\cal F}\star ...\star {\cal F})$,
the reason being that such 
forms are not closed nor invariant as one easily sees by applying
cyclic equivalence \footnote{One may think to get
around this obstacle by allowing for cyclic equivalence in each 
trace separately, which amounts to changing the cohomology. In this case one
would get a closed and invariant top form but such cyclic equivalence would
clash with the integration rule in D dimensions, so that the resulting
integral
 $\int d^Dx\Omega_D^1$ would not satisfy the Wess-Zumino consistency
conditions.}.

A similar argument shows that in \nc gauge theories there cannot be
mixed anomalies \cite{Mar,Int}. For suppose we have a bifundamental gauge
theory
 with gauge group $U(N_1)\times U(N_2)$. In this theory we can have 
$U(N_1)$ anomalies and $U(N_2)$ anomalies, but not mixed
$U(N_1)\times U(N_2)$ anomalies. The reason is that the latter kind
of anomalies should come for instance from a top form like
$\Tr_1({\cal F}_1\star ...\star {\cal F}_1)
\Tr_2({\cal F}_2\star ...\star {\cal F}_2)$, 
where the first trace refers
to $U(N_1)$ and the second to $U(N_2)$. For the same reason as before
such top form is neither closed nor invariant.

Finally we would like to make a comment on the $U(1)$ anomaly inside
a $U(N)$ theory. From what has been just said it is apparent that
there is no room for a separate $U(1)$ anomaly (differently from 
ordinary $U(N)$ gauge theories). Anyhow one can verify it
directly. For example, in the simplest case, $n=1 \,(D=2)$, we can try
to split $h^{AB}= \frac 12 \delta^{AB}$ into a $U(1)$ part $\delta^{00}$
and an $SU(N)$ part $\delta^{ab}$ and build two corresponding separate
top forms. However it is easy to see that $\delta^{00}$ and 
$\delta^{ab}$ do not satisfy separately (\ref{hconstr}). This can be
extended to higher dimensional cases and is of course in keeping 
with the impossibility to disentangle the $U(1)$ factor in a
\nc $U(N)$ gauge theory.

\section{Anomalies in the adjoint representation}
 
In ordinary gauge theories with chiral fermions in the adjoint 
representation the chiral anomaly identically vanishes in $D=4k$
while it is nonvanishing in $D=4k+2$ dimensions. In ordinary gauge
theories in $D=4k$ one can verify this by a direct Feynman diagram 
computation. Or else one can get the explicit form of the adjoint
anomaly via descent equations starting from the top form with ${\cal F}$
valued in the adjoint representation,
 i.e. expanded over the set of
antisymmetric matrices $F^A$ (\ref{FD}). In this way one sees that,
in $D=4k$ dimensions, the top form (and consequently the anomaly)
is determined by the trace of the  symmetric product of 
$2k+1$
antisymmetric $F^A$ matrices. Therefore
it identically vanishes.

In \nc gauge theories the question of chiral anomaly in the adjoint
representation looks at first a bit puzzling. Let us see why.
In \cite{Mar} the adjoint chiral anomaly has been shown to vanish 
in $D=4$ by writing the action in terms of Majorana fermions and showing the
vector nature of the vertex, and
by a direct Feynman diagram calculation. On the other hand  if we try to apply
to this case the formula obtained from the descent
 equations we see
immediately that in $D=4$ we will never get zero, 
 the reason being that
$Tr(X^AX^BX^C)$ cannot vanish for any of the
 representations considered in
the previous section, nor do we know
 of other representations of the group of
gauge transformations whose
 generators, when inserted in the above trace, can
give zero. 
 So it is evident that for the adjoint representation the formulas

 obtained via the descent equations must be applied with a grain of salt.

To clarify the situation let us start from the relevant \nc action 
\be
S = \int d^Dx\, \bar \psi^j{}_i \star 
\gamma^\mu(i \d_\mu \psi^i{}_j
+ A_\mu^i{}_k\star P_+ \psi^k{}_j - 
P_+\psi^i{}_k\star A_\mu^k{}_j).\label{action}
\ee
We find it useful to rewrite this action in terms of the basis 
introduced in section 2: $\psi^j{}_i= \psi_A (t^A)^j{}_i$ and so on.
The action (\ref{action}) takes the form
\be
S = \int d^Dx \,\bar \psi\star \gamma^\mu(i \d_\mu \psi
+ \widehat A_\mu \star P_+\psi - P_+ \psi\star \widehat A_\mu),
\label{action'}
\ee
where we use a vector notation for $\psi= \{\psi^A\}$ and a matrix
notation for the gauge potential, $\widehat A_\mu = A_\mu^B G^B$. 
In particular the last term in (\ref{action'}) means
\be
\bar \psi\star \gamma^\mu P_+\psi\star \widehat A_\mu=
\bar \psi^B\star \gamma^\mu P_+\psi^C\star (\widehat A_\mu)_{CB}\0
\ee
This action is invariant under
\be
&&\delta \widehat A_\mu = \d_\mu\widehat \lambda -i \widehat A_\mu \star\widehat \lambda
+ i \widehat \lambda\star \widehat A_\mu\0\\
&&\delta \psi = i \widehat \lambda \star\psi - i \psi\star \widehat \lambda\0\\
&& \delta \bar \psi =- i \widehat \lambda \star\bar\psi + i 
\bar\psi \star\widehat \lambda\0 
\ee
and, again, $\widehat \lambda = \lambda^B G^B$. 

Now let us introduce the charge conjugate field 
$\psi^c= C^\dagger \bar\psi^T$. The charge conjugation operator $C$
is defined to have the following properties: 
\be
C^\dagger C =1, \quad\quad C\gamma_\mu C^\dagger = - \gamma_\mu^T
\label{C}
\ee
Moreover we assume a metric $g_{\mu\nu}$ with signature 
$(+,-,\ldots,-)$ and $\gamma_0^*=-\gamma_0$. As a consequence,
in dimension $D=2n$ we have 
\be
C \widehat \gamma C^\dagger = (-1)^{n}\widehat\gamma.\label{gamma}
\ee
where $\widehat\gamma$ has been defined in section 2.
If we express the action (\ref{action'}) in terms of the charge--conjugate
fields we get
\be
S = \int d^Dx \, \bar \psi^c\star \gamma^\mu(i \d_\mu \psi^c 
+ \widehat A_\mu \star P'_+\psi^c - P'_+ \psi^c\star \widehat A_\mu) 
\label{action''} 
\ee 
where $P_+'= P_-$ in dimension $D=4k$ and $P_+'= P_+$ in dimension $D=4k+2$.
Since integrating over $\psi$ or $\psi^c$ in the path integral does not
entail any difference, we see that, when computing the anomaly by Feynman
diagrams techniques, (i) in dimension $D=4k$, the action (\ref{action'}) and
the action (\ref{action''}) give opposite contributions to the anomaly
because they contain fermions with opposite chirality, therefore 
the anomaly must vanish, (ii) in dimension $D=4k+2$, the action
(\ref{action'})  and the action (\ref{action''}) give the same contribution,
therefore the anomaly presumably will not vanish.

This clarifies the problem in $D=4k$ dimensions. What remains for us to do
is to compute the anomaly in $D=4k+2$. The solution is actually very simple:
from section 3 we learned that there is only one nontrivial cocycle
in any even $D=2n$ dimension, it is the one determined by the top form
$\widehat{\tr}(\widehat F\star\ldots \star\widehat F)$ with $n+1$ entries. The
corresponding (unintegrated) anomaly is given  by formula (\ref{anom}). 

Of course what remains to be determined is the coefficient in front
of the anomaly for the present case.
To determine it we must resort to other methods. 
Here we follow \cite{Mar} and
express the gauge current in (\ref{action'}) as the sum of two pieces \be
&&j^B_\mu = j^{+B}_\mu + j^{-B}_\mu,\label{curr}\\ 
&&j^{+B}_\mu = \bar \psi \star G^B \gamma_\mu P_+\psi,\0\\
&& j^{-B}_\mu = -\bar \psi \star\gamma_\mu P_+\psi G^B=
\bar \psi^c \star G^B \gamma_\mu P_+\psi^c \0
\ee
Since the two pieces represent the same vertex (replacing $\psi$ by
$\psi^c$) they must contribute the same amount to the anomaly, so, in
fact, it is enough to compute the contribution from one, $j^+$ for instance.

Let us solve now an auxiliary problem. We remark that
$j^+$ specifies the fermion-gluon interaction  corresponding to the
following action \be 
S = \int d^Dx \, \bar \psi\star \gamma^\mu(i \d_\mu \psi 
+ \widehat A_\mu \star P_+\psi) 
\label{act} 
\ee 
which looks like (\ref{actfund}), except that instead of the generators 
$t^A$ we have here the generators $G^A$. Therefore this action is
invariant under
\be 
&&\delta \widehat A_\mu = \d_\mu\widehat \lambda -i \widehat A_\mu \star\widehat \lambda 
+ i \widehat \lambda\star \widehat A_\mu\0\\ 
&&\delta \psi = i \widehat \lambda \star\psi, \quad\quad 
\delta \bar \psi = - i\bar\psi \star\widehat \lambda.\0  
\ee 
We know how to compute the anomaly of (\ref{act}) in any even dimension. 
Simply we apply the descent equations method to the closed invariant form 
$\Omega_{2n+2}= \widehat{\tr} (\widehat F \star \widehat F...\star \widehat F)$ with
$n+1$  $\widehat F$
entries, where $\widehat F$ is the curvature of $\widehat A$. From (\ref{trNTR})
we know that $\widehat{\tr} (\widehat F \star \widehat F...\star \widehat F)=
N  \tr(F\star F \star ...\star F)$ where the RHS refers to the fundamental
representation. Therefore the corresponding anomaly is $N$ times the
anomaly in the fundamental representation. As far as the adjoint anomaly
is concerned, we would be therefore led to conclude that this anomaly 
cancels against the
corresponding negative contribution from $j^-$ in $D=4k$, while it gets
doubled in $D=4k+2$. Therefore our conclusion would be that in $D=4k+2$
dimensions the chiral adjoint anomaly is $2N$ times the chiral anomaly 
in the fundamental representation.

However we are not quite finished yet. 
One may rightly object that above we did not really compute the anomaly of 
(\ref{action}) or (\ref{action'}), but only twice the anomaly of (\ref{act}).
We have still to prove that there are no interference terms between
$j^+$ and $j^-$. The lowest order contribution to the anomaly in $D=4k+2$
comes from the $(k+2)$-point function of $j^+$ and the $(k+2)$-point function
of $j^-$. They are equal and proportional to
\be
\int d^Dx \widehat{\tr}(\widehat C\star d\widehat A \star\ldots \star d\widehat A)\0
\ee
with $k+1$ $d\widehat A$ entries. This is exactly the first term of the unique
cocycle appropriate for $D=4k+2$ and confirms what we said above.
There are however other possible contributions which may come from mixed
correlators with both $j^+$ and $j^-$ entries. The latter are proportional
to traces of the type $\widehat{\tr} (G^{A_1}\ldots (G^{B_1})^T\dots)$ in
which some of the entries are transposed matrices. Thanks to eq.(\ref{comm})
we can regroup all the transposed matrices on the right. Then applying
repeatedly formulas (\ref{FD}) as well as (\ref{commFD}) it
is easy to see that these traces are reducible. For instance
\be
\widehat{\tr} \left(G^AG^B (G^C)^T (G^D)^T\right) = 
\frac {1}{N^2} \widehat{\tr} (G^AG^B)
\widehat{\tr}(G^CG^D).\0
\ee
But we know that this cannot correspond to any BRST cocycle. Therefore
mixed correlators  with both $j^+$ and $j^-$ entries cannot contribute to 
the anomaly. Therefore our conclusion is that in $D=4k+2$
dimensions the chiral adjoint anomaly is $2N$ times the chiral anomaly 
in the fundamental representation.

The uniqueness of the anomaly cocycle is a distinctive element of \nc
chiral gauge theories as compared the ordinary ones.  
To see this let us take as an example $D=6$ and let us analyse the distinct
nontrivial ordinary cocycles contained in the only \nc one.
The latter is determined by $h^{ABCD}=\tr (t^At^Bt^Ct^D)$. It should be
noticed that when dealing with the ordinary $U(N)$ gauge theory only
the completely symmetric part $h^{(ABCD)}$ of $h^{ABCD}$ is relevant, as 
the other parts of the cocycle vanish. It is then easy to list
the independent ordinary cocycles. They are determined by:
\begin{itemize}
\item $h^{0000}$ , which gives rise to the pure $U(1)$ anomaly;
\item $h^{(0abc)}\sim d^{abc}$, which gives rise to a mixed 
$U(1)\times SU(N)$ anomaly;
\item $h^{(00ab)}\sim \delta^{ab}$ which gives rise to another mixed
$U(1)\times SU(N)$  anomaly;
\item $h^{(abcd)}$ which is a combination of $\delta^{(ab} \delta^{cd)}$
and of $d^{(abx} d_x{}^{cd)}$; the first gives rise to a reducible 
$SU(N)$ anomaly, the second to the irreducible $SU(N)$ anomaly.
\end{itemize}
All these independent ordinary cocycles coalesce to form a unique
\nc cocycle. Finally, it is interesting to remark that the ratio between 
the irreducible ordinary $SU(N)$
anomaly in the adjoint and in the fundamental representation is again $2N$.

\section{Conclusion}

In this paper, using the principle of nc--locality, we have calculated
the anomalies in a chiral \nc $U(N)$ gauge theories for all admissible
representations of the group of gauge transformations in any even
dimension. We have shown in particular that, differently from ordinary
theories, 
\begin{itemize}
\item there do not exist reducible anomalies; 
\item there do not exist mixed anomalies;
\item there exist only one possible anomaly for each even dimension.
\end{itemize}
Moreover we have explicitly calculated the chiral anomaly in the adjoint
representation and found that
\begin{itemize}
\item it vanishes in $D=4k$, as in ordinary theories;
\item in $D=4k+2$ it equals $2N$ times the anomaly in the fundamental
representation; in ordinary theories this property holds generally only
for the irreducible part.
\end{itemize}

We add a short comment concerning covariant anomalies. Covariant anomalies
are particular cases of consistent ones. They can be calculated from
a nonchiral (i.e. with Dirac fermions) gauge theory by coupling it
to a vector plus a pseudovector potential $V_\mu +\widehat \gamma A_\mu$,
and computing the anomaly of the chiral
current,\cite{Bardeen}. This anomaly satisfies WZ
consistency conditions (more complicated than the above ones, see 
for instance \cite{AB}). The covariant anomaly is obtained by 
eventually setting $A_\mu=0$. There is therefore a one-to-one correspondence
between covariant and consistent anomalies, they are determined
by the same (unsymmetrized) traces of generators in the appropriate
representations and the former (in a vector theory) vanish when and only 
when the latter (in the corresponding chiral theory) do.

Finally the implications  of the results recently obtained for anomaly 
calculations on the lattice in \cite{Nish} are not clear to us.

\vskip .5cm
{\bf Acknowledgments}
A.S. would like to thank SISSA for the 
hospitality extended to him during this research. L.B. would like to thank
M.Henneaux and A.Schwimmer for an exchange of e-mail messages.
This research was partially supported by the Italian MIUR under the 
program ``Fisica Teorica delle Interazioni Fondamentali'' and  by
the RFBR Grant no. 99-02-18417.

\end{document}